\documentclass[prb,nopacs,twocolumn,preprintnumbers,amsmath,amssymb]{revtex4}

\usepackage{epsfig}
\usepackage{graphicx}
\usepackage{dcolumn}
\usepackage{color}
\usepackage{enumerate}
\usepackage{bm}
\newcommand{\be}{\begin{equation}}
\newcommand{\ee}{\end{equation}}

\begin{document}

\title{Topological spin-transfer drag driven by skyrmion diffusion}

\author{H\'ector Ochoa, Se Kwon Kim, and Yaroslav Tserkovnyak}

{\affiliation{Department of Physics and Astronomy, University of California, Los Angeles, California 90095, USA}

\begin{abstract}
We study the spin-transfer drag mediated by the Brownian motion of skyrmions. The essential idea is illustrated in a two-terminal geometry, in which a thin film of a magnetic insulator is placed in between two metallic reservoirs. An electric current in one of the terminals pumps topological charge into the magnet via a spin-transfer torque. The charge diffuses over the bulk of the system as stable skyrmion textures. By Onsager's reciprocity, the topological charge leaving the magnet produces an electromotive force in the second terminal. The voltage signal decays algebraically with the separation between contacts, in contrast to the exponential suppression of the spin drag driven by non-protected excitations like magnons. We show how this topological effect can be used as a tool to characterize the phase diagram of chiral magnets and thin films with interfacial Dzyaloshinskii-Moriya interactions.
\end{abstract}

\maketitle

\section{Introduction}

\label{sec:Intro}

Magnetic insulators stand out as promising platforms for spintronics devices due to the lack of energy dissipation by Joule heating. Nevertheless, the transmission of information encoded in the collective dynamics of localized spins is not immune to losses due to the exchange of angular momentum with itinerant electrons and the lattice. In condensed matter systems, dissipationless transport is either sustained by a superfluid ground state or driven by topological excitations. Spin superfluidity has been extensively discussed in the context of easy-plane magnetic insulators.\cite{spin-superfluidity} Long-ranged spin transmission is supported by the coherent precession of the order parameter within the easy-plane of the magnet, which, on the other hand, is not robust under perturbations breaking the U(1) spin symmetry. Dissipationless spin transport can be mediated also by the Brownian motion of solitons like, for example, domain walls.\cite{DW_superfluidity} In that case, however, thermally activated phase slip events\cite{phase_slips} invalidate the topological protection of domain walls' chirality, imposing restrictions on the geometry of the device. The aim of this work is to generalize the idea of spin transport mediated by solitons, focusing on skyrmion textures in order to overcome these limitations.

Magnetic skyrmions are characterized by a topological index that labels the number of times that the local order parameter wraps the unit sphere in spin space. This integer --the skyrmion charge-- remains unchanged as long as the texture varies slowly. Due to its robustness, skyrmions are promising candidates as building blocks for information storage.\cite{rev} The observation of skyrmions in bulk\cite{MnSi,Cu2OSeO3} and thin films\cite{MnSi_film,Cu2OSeO3_film} of chiral magnets, or in systems with interfacial Dzyaloshinskii-Moriya interaction,\cite{science} together with the low spin-polarized currents that are needed to move them\cite{current} has boosted the field in the recent years.

\begin{figure}
\includegraphics[width=1\columnwidth]{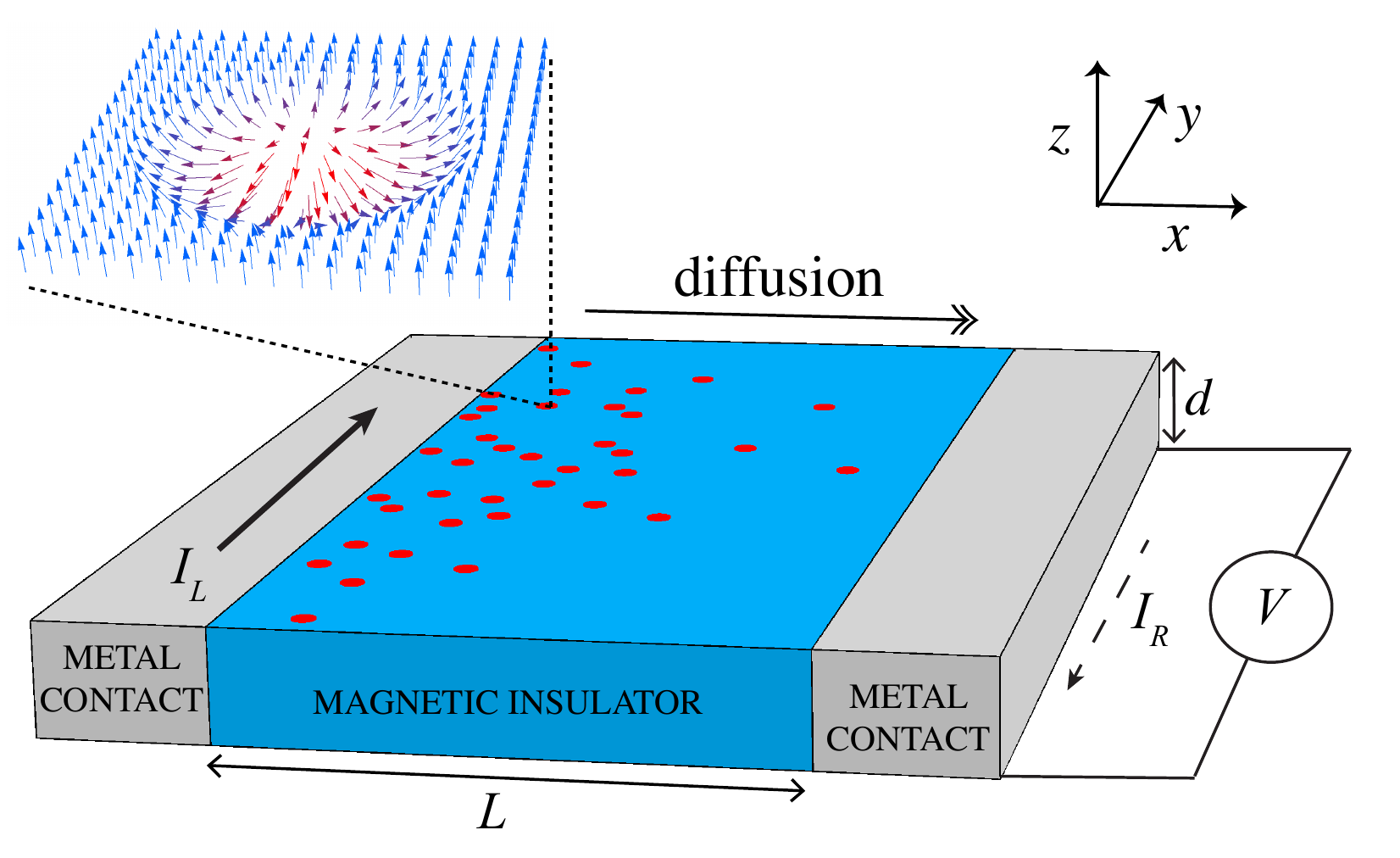}
\caption{Scheme for electrical injection and detection of skyrmions. The electric current in the left terminal pumps skyrmion charge into the magnet, which diffuses as stable solitons over the system. The skyrmion charge leaving the system sustains a voltage signal in the second terminal.}
\label{fig:geometry}
\end{figure}

Let us consider the two-terminal geometry depicted in Fig.~\ref{fig:geometry}. A current in the left contact exerts a torque on the order parameter of the film, favoring the injection of skyrmion charge. The charge is topologically protected, so it diffuses without losses over the bulk of the system as stable skyrmion solitons, which eventually reach the right terminal. By the reciprocal effect to the spin-transfer torque, the topological charge leaving the system pumps itinerant spins into the right metal, generating an electromotive force. The drag of spin current is negative, contrary to frictional effects based on linear momentum transfer. In the steady state and neglecting boundary effects, the drag coefficient $\mathcal{C}_{d}\equiv I_R/I_L$ reduces to
\begin{align}
\label{eq:Cd}
\mathcal{C}_d=-\mu\rho_{0}\sigma\left(\frac{2\pi\hbar\mathcal{P}}{e}\right)^2\frac{d}{L}.
\end{align}
Here $\rho_{0}$ is the concentration of skyrmions at the equilibrium, $\mu$ is the longitudinal skyrmion mobility, and $\sigma$ is the conductivity of the metal contacts. The term between brackets must be interpreted as the conversion factor between charge and spin current, where $\mathcal{P}$ is a dimensionless parameters measuring the efficiency of this conversion. The last factor is geometrical, $d$ and $L$ being the thickness of the film and the distance between terminals, respectively. Its origin is the following: on the one hand, the spin transfer and pumping effects are more efficient as the surface of the interface grows; on the other, the drag effect decays algebraically with the distance between contacts due to the diffusion of the skyrmion charge. The latter is a direct consequence of the topological protection of carriers, in contrast to the exponential decay of the spin signal mediated by thermal magnons.\cite{magnon_transport}

For the rest of the manuscript, we formalize these ideas and derive the result disclosed in Eq.~\eqref{eq:Cd}. In Sec.~\ref{sec:dynamics}, we write a continuity equation for the skyrmion charge density and show how non-equilibrium torques at the interface with metals pump skyrmion charge into the magnet, and reciprocally, skyrmion charge leaving the magnet pumps itinerant spins in the metal. Then, in Sec.~\ref{sec:gas}, we complete the hydrodynamic picture for skyrmion charge nucleation and diffusion, what allows us to write the spin drag coefficient in terms of microscopic parameters of the system in Sec.~\ref{sec:drag}. We discuss the scope of our results in Sec.~\ref{sec:discussion}.

\section{General dynamics}

\label{sec:dynamics}

We consider a two-dimensional magnetic insulator at temperatures well below the ordering transition. We are primarily interested in the dynamics of $\mathbf{n}\left(\mathbf{r}\right)$, which is defined as the unit vector along the local spin order parameter, $\mathbf{s}\left(\mathbf{r}\right)=s\mathbf{n}\left(\mathbf{r}\right)$. The modulus $s\equiv\left|\mathbf{s}\left(\mathbf{r}\right)\right|$ (with units of angular momentum per area) is assumed to be fixed at its saturation value. The dynamics of $\mathbf{n}\left(\mathbf{r}\right)$ is governed by the Landau-Lifshitz-Gilbert equation\cite{LLG}
\begin{align}
s\left(1+\alpha\mathbf{n}\times\right)\dot{\mathbf{n}}=\mathbf{n}\times\mathbf{h}_{eff}+\boldsymbol{\tau}.
\label{eq:S-LLG}
\end{align}
Here $\mathbf{h}_{eff}$ is the effective exchange field derived from the magnetic free energy functional of the system,
\begin{align*}
\mathbf{h}_{eff}\equiv-\frac{\delta \mathcal{U}\left[\mathbf{n}\left(\mathbf{r}\right)\right]}{\delta\mathbf{n}}.
\end{align*}
The phenomenological parameter $\alpha$ accounts for Gilbert damping, which measures the rate of energy dissipation through a Rayleigh function of the form\begin{align}
\label{eq:dissipation}
\mathcal{R}\left[\dot{\mathbf{n}}\right]=\frac{\alpha s}{2}\int d^2\mathbf{r}\text{ }\dot{\mathbf{n}}^2.
\end{align}

In the interface with metal contacts, we must consider also the action of non-equilibrium spin-transfer torques, the last term in the right hand of Eq.~\eqref{eq:S-LLG}. An electric current in the left contact exerts a reactive torque on the order parameter of the film, which, to the lowest order in spatial gradients of $\mathbf{n}\left(\mathbf{r}\right)$, adopts the following form:\cite{Berger}
\begin{align}
\label{eq:Berger}
\boldsymbol{\tau}_L=\frac{\hbar}{2e}\mathcal{P}\left(\vec{\mathcal{J}}_L\cdot\vec{\nabla}\right)\mathbf{n}.
\end{align}
Here $\vec{\mathcal{J}}_L$ is the current (linear) density along the interface and $\mathcal{P}$ is a dimensionless phenomenological parameter. The torque in Eq.~\eqref{eq:Berger} appears as the leading contribution in the limit of a strong exchange field created in the metal by proximity with the magnet. In such a regime, the itinerant spins follow adiabatically the dynamics of the order parameter of the film.\cite{pumping} The same expression would be obtained in the case of a ferromagnetic contact, $\mathcal{P}$ being a measure of the degree of spin polarization of the current in that case.

\subsection{Continuity equation}

\label{subsec:bulk}

The order parameter $\mathbf{n}\left(\mathbf{r}\right)$ defines a mapping between the coordinate space, $\mathbf{r}=\left(x,y\right)$, and the unit sphere in spin space. We consider a finite energy excitation on top of the uniformly ordered ground state. The uniform background imposes boundary conditions on $\mathbf{n}\left(\mathbf{r}\right)$, in such a way that the coordinate space is compactificated into a sphere. Consequently, $\mathbf{n}\left(\mathbf{r}\right)$ can be classified according to the second homotopy group $\pi_2\left(S^2\right)=\mathbb{Z}$. The corresponding integer index is the skyrmion charge,
\begin{align}
\mathcal{Q}\equiv-\frac{1}{4\pi}\int d^2\mathbf{r}\text{ }\mathbf{n}\cdot\left(\partial_x\mathbf{n}\times\partial_y\mathbf{n}\right).
\label{eq:skyrmion_charge}
\end{align}

Magnetic textures with different $\mathcal{Q}$ are not smoothly connected to each other and belong to different energy sectors,\cite{Belavin_Polyakov} separated by energy barriers that we assume to be much larger than the temperature of the system. The latter condition prevents the texture to slip into an atomic size defect and guarantees its topological protection. This must be interpreted as the conservation of $\mathcal{Q}$, which can be expressed as a continuity equation for the skyrmion charge density, $\rho\equiv -\frac{1}{4\pi}\mathbf{n}\cdot\left(\partial_x\mathbf{n}\times\partial_y\mathbf{n}\right)$. We can write then\cite{book}
\begin{align}
\partial_t\rho+\vec{\nabla}\cdot\vec{j}=0,
\label{eq:continuity}
\end{align}
where the skyrmion charge current reads as
\begin{align}
\vec{j}\equiv\frac{1}{4\pi}\mathbf{n}\cdot\left[\left(\hat{z}\times\vec{\nabla}\right)\mathbf{n}\times\dot{\mathbf{n}}\right].
\label{eq:sky_current}
\end{align}
Eq.~\eqref{eq:continuity} follows immediately from this definition as long as $\mathbf{n}\left(\mathbf{r}\right)$ does not contain singular points.

\subsection{Skyrmion charge pumping}

\begin{figure*}
\includegraphics[width=\textwidth]{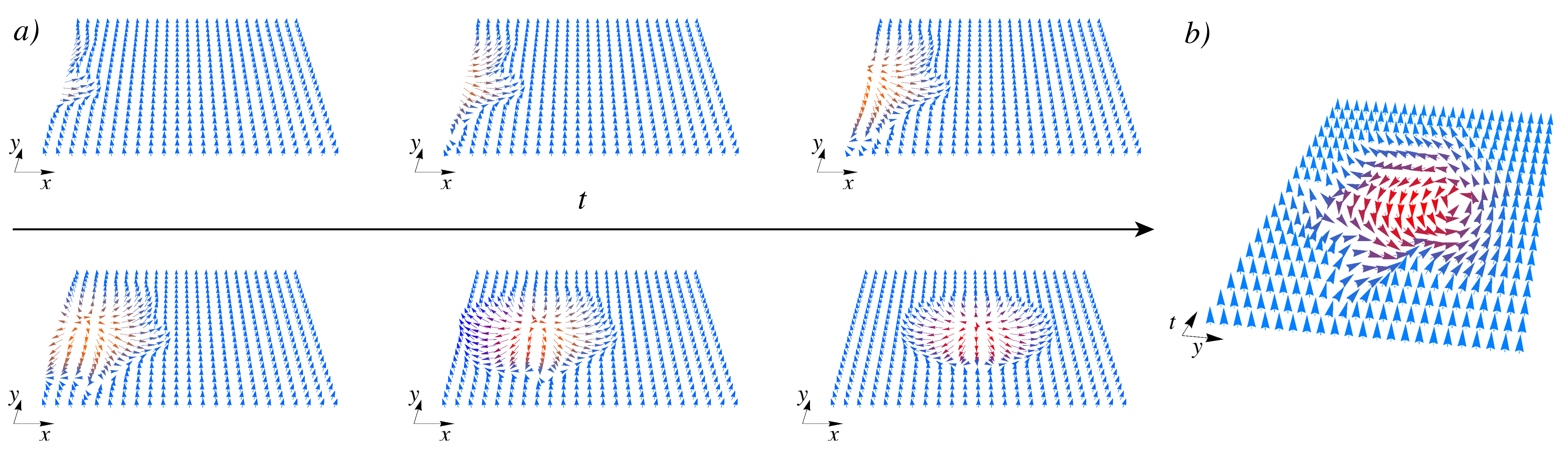}
\caption{a) Sketch of the injection of a skyrmion through one of the leads. The non-equilibrium torque favors the nucleation of skyrmion charge, which diffuses into the bulk of the film as a metastable soliton. b) Dynamics of the localized spins at the edge. According to Eqs.~\eqref{eq:continuity}-\eqref{eq:sky_current}, the skyrmion number of the texture in the $(y,t)$ plane equals the skyrmion charge of the injected soliton.}
\label{fig:sky}
\end{figure*}

Let us consider the work carried out by the spin-transfer torque,
\begin{align*}
W=\int_{S} \delta\mathbf{n}\cdot\mathbf{h}_{\tau},
\end{align*}
where $\delta\mathbf{n}$ expresses the change of the order parameter at the interface, $\mathbf{h}_{\tau}\equiv\boldsymbol{\tau}_L\times\mathbf{n}$ is the effective field associated with the spin-transfer torque, and the integral is performed along the interface $S$ between the contact and the film. The work per unit area can be evaluated as
\begin{align*}
w=\delta\mathbf{n}\cdot\boldsymbol{\tau}_{L}\times\mathbf{n}=\int dt\text{ }\dot{\mathbf{n}}\cdot\boldsymbol{\tau}_L\times\mathbf{n}.
\label{eq:w}
\end{align*}
Then, plugging  Eq.~\eqref{eq:Berger} into this expression leads to
\begin{gather}
w=\frac{\hbar}{2e}\mathcal{P}\int dt\text{ }\mathbf{n}\cdot\left[\dot{\mathbf{n}}\times\left(\vec{\mathcal{J}}_L\cdot\vec{\nabla}\right)\mathbf{n}\right]
\nonumber\\
=\frac{2\pi\hbar\mathcal{P}}{e}\int dt\text{ }\hat{z}\cdot\left(\vec{j}\times\vec{\mathcal{J}}_L\right),
\end{gather}
where the second line follows immediately from the definition of the skyrmion charge current  in Eq.~\eqref{eq:sky_current}. We have to perform the remaining integral along the interface; the final result reads\begin{align}
W=\frac{2\pi\hbar\mathcal{P}}{e}I_L\int dt \int dy\text{ }j_{x}=\frac{2\pi\hbar\mathcal{P}}{e}I_L\mathcal{Q},
\label{eq:W}
\end{align}
where $I_L=d\mathcal{J}_L^y$. Notice that the integral of $j_{x}$ equals the skyrmion charge injected into the system through the interface, as implied by the continuity equation~\eqref{eq:continuity}.

As deduced from Eq.~\eqref{eq:W}, the spin-transfer torque works in favor of the formation of skyrmion textures with $\mathcal{Q}>0$, injecting topological charge into the magnetic system. The process is depicted in Fig.~\ref{fig:sky}~a). The charge of the injected skyrmion equals the skyrmion number of the texture in the $\left(y,t\right)$ plane, Fig.~\ref{fig:sky}~b). The quantization of $W$ in units of $\hbar I/e$ resembles the quantization of particle transport in the theory of adiabatic pumping\cite{Thouless} if we identify the period of the driving potential in that case with the characteristic time for the injection of a skyrmion. This analogy breaks down in the presence of spin realaxation at the interface. On symmetry grounds, it is possible to construct another spin-transfer torque to the same order in spatial gradients of the form\cite{Zhang_Li}
\begin{align}
\label{eq:beta}
\boldsymbol{\tau}=\frac{\hbar\beta}{2e}\text{ }\mathcal{P}\text{ }\mathbf{n}\times\left(\vec{\mathcal{J}}_L\cdot\vec{\nabla}\right)\mathbf{n}.
\end{align}
Notice that this torque is dissipative (odd under inversion of time), contrary to the one in Eq.~\eqref{eq:Berger}. Here $\beta$ is a dimensionless parameter that accounts for spin relaxation. Microscopic calculations\cite{micro} reveal that $\beta=\hbar/\tau_s\Delta_{ex}$, $\tau_s$ and $\Delta_{ex}$ being the spin relaxation time and the proximity exchange coupling constant at the interface, respectively. The work carried out by the torque in Eq.~\eqref{eq:beta} cannot be expressed in terms of the skyrmion charge current anymore. Only when the itinerant spins follow adiabatically the dynamics of the order parameter, the exchange of angular momentum can be identified with the pumping of geometrical phase from one system to the other. Such an adiabatic regime corresponds then to the limit of a strong exchange coupling, $\Delta_{ex}\gg\hbar\tau_s^{-1}$.

\subsection{Voltage signal due to skyrmion dynamics}

Just as a spin-polarized current exerts a torque on the order parameter of the film, a time-dependent magnetic texture pumps spins into the metal. The one is the reciprocal process of the other in Onsager's sense. The dynamics of the order parameter at the right terminal induces an electromotive force of the form\cite{pumping}
\begin{align}
\vec{\mathcal{E}}_R=\frac{\hbar}{2e}\mathcal{P}\text{ }\mathbf{n}\cdot\left(\vec{\nabla}\mathbf{n} \times \dot{\mathbf{n}}\right).
\end{align}
This expression is valid in the adiabatic limit, otherwise an additional term proportional to $\beta$ must be considered.

From Eq.~\eqref{eq:sky_current}, we can rewrite the electromotive force in terms of the skyrmion charge current as\begin{align}
\vec{\mathcal{E}}_R=-\frac{2\pi\hbar\mathcal{P}}{e}\text{}\hat{z}\times\vec{j},
\end{align}
and therefore the voltage in the second terminal is just \begin{align}
V=\int d\vec{\ell} \cdot\vec{\mathcal{E}}_R=-\frac{2\pi\hbar\mathcal{P}}{e}\int dy \text{ }j_x=\frac{2\pi\hbar\mathcal{P}}{e}\dot{\mathcal{Q}},
\label{eq:EMF}
\end{align}
where $\dot{\mathcal{Q}}<0$ is the rate of skyrmion charge leaving the system.

\section{Skyrmion gas}

\label{sec:gas}

From now on, we assume that the thin film hosts stable skyrmion textures, particle-like solutions of energy $E_0$ and charge $\mathcal{Q}$ in the bulk of the system. These rigid textures must be interpreted as the carriers of the skyrmion charge density, completing in this way the hydrodynamic picture provided by Eq.~\eqref{eq:continuity}.

\subsection{Diffusion}

\label{subsec:diff}

The existence of rigid skyrmion textures implies that the dynamics of the order parameter can described in terms of \textit{hard} and \textit{soft} modes. The later is identified with the center of mass of the skyrmion, $\vec{\mathcal{R}}$, whereas the former are modes that relax very fast, in a microscopic, short time scale. This microscopic relaxation time characterizes the approach to a local equilibrium state, determined by the local values of the conserved densities. The condition of local equilibrium implies that $\vec{j}$ can be written as a functional of $\rho$, and therefore it admits an expansion of the form
\begin{align}
\vec{j}\left(\mathbf{r},t\right)=\vec{j}\left[\rho\left(\mathbf{r},t\right)\right]\approx\rho\text{ }\vec{v}-\hat{D}\cdot\vec{\nabla}\rho.
\label{eq:sky_current_hydrodynamic}
\end{align}
Here $\vec{v}=\dot{\vec{\mathcal{R}}}$ is the velocity of skyrmions and $\hat{D}$ is the diffusion tensor.

At thermodynamic equilibrium we have\begin{align*}
\rho_0\propto e^{-\frac{\mathcal{U}\left[\mathbf{n}\left(\mathbf{r}\right)\right]}{k_BT}}\left(\equiv e^{-\frac{E_0}{k_BT}}\right),
\end{align*}
and therefore\begin{align*}
\vec{\nabla}\rho_0=-\frac{\vec{\nabla}\mathcal{U}}{k_BT}\rho_0=\frac{\vec{F}}{k_BT}\rho_0.
\end{align*}
The force acting on the skyrmion $\vec{F}$ is related with its velocity through the constitutive relation\begin{align}
\vec{v}=\hat{\mu}\cdot\vec{F},
\end{align}
where $\hat{\mu}$ is the mobility tensor. From these relations and the condition of thermodynamic equilibrium, $\vec{j}=0$, we deduce the Einstein-Smoluchowski identity:\begin{align}
\label{eq:ES}
\hat{D}=k_BT\hat{\mu}.
\end{align}

The dynamical equation for the center of mass of the skyrmion can be derived from Eq.~\eqref{eq:S-LLG} within a collective coordinate approach;\cite{DW_dynamics} we obtain
\begin{align}
\label{eq:Langevin}
\hat{\eta}\cdot\vec{v}+4\pi s\mathcal{Q}\text{ }\hat{z}\times\vec{v}=\vec{F}.
\end{align}
The first term is associated with Gilbert damping, $\hat{\eta}$ being a generalized viscosity tensor whose components read\begin{align}
\eta_{ij}=\alpha s\int d^2\mathbf{r}\text{ }\partial_i\mathbf{n}\cdot\partial_j\mathbf{n}.
\end{align}
The second term is the Magnus force. From Eq.~\eqref{eq:Langevin} we identify the mobility tensor as\begin{align}
\hat{\mu}=\left(\hat{\eta}+\hat{G}\right)^{-1},
\label{eq:mu}
\end{align}
where $G_{ij}=4\pi s\mathcal{Q}\epsilon_{ji}$ is the gyromagnetic tensor.

Notice that the mobility tensor splits in two contributions with very different physical meaning. The symmetric component is dissipative and can be derived from the Fokker-Planck equation describing the stochastic dynamics of skyrmions.\cite{FP} The antisymmetric component is non-dissipative and proportional to $\mathcal{Q}$. Its origin can be understood in analogy with the dynamics of a charged particle in a magnetic field.\cite{Tchernyshyov} If we assume translational invariance along $\hat{y}$ in the geometry of Fig.~\eqref{fig:geometry}, then only the dissipative term contributes to the diffusive skyrmion current leaving the system, $j_x=-D_{xx}\partial_x\rho$, where\begin{align}
D_{xx}\equiv D=\frac{k_BT\eta_{yy}}{\eta_{xx}\eta_{yy}-\eta_{xy}^2+16\pi^2s^2\mathcal{Q}^2}.
\end{align}

\subsection{Nucleation theory}

\label{subsec:nuc}

Next, we evaluate the rate of skyrmion charge pumping. In the spirit of the reaction-rate theory,\cite{reaction-rate} we can write the skyrmion charge current at the boundaries of the film as
\begin{align}
j_{x}\left(I\right)=\gamma \left(T,I\right)-\bar{\gamma}\left(T\right)\rho.
\label{eq:reaction-rate}
\end{align}
Here $\gamma\left(T,I\right)$ represents the nucleation rate and $\bar{\gamma}\left(T\right)$ is the annihilation rate per unit density. The condition $j_{x}\left(I=0\right)=0$ defines the skyrmion charge density at equilibrium,
\begin{align}
\rho_0\equiv\frac{\gamma\left(T\right)}{\bar{\gamma}\left(T\right)}\propto e^{-\frac{E_0}{k_BT}}.
\label{eq:rho_0}
\end{align}

The annihilation rate can be interpreted as the characteristic escape velocity of skyrmions, which is a bulk property and therefore is expected not to depend on the current at the leads. For the nucleation rate, we can write in general
\begin{align}
\gamma\left(T,I\right)=\nu\left(T\right) e^{-\frac{E}{k_BT}},
\end{align}
where $E$ is the energy barrier for the nucleation of a skyrmion and $\nu\left(T\right)$ represents the \textit{attempt frequency}. The latter does not depend on the current to the leading order,\cite{attempt} whereas the former can be written as $E=E_0-W$, where $W$ is the work carried out by the spin-transfer toque, Eq.~\eqref{eq:W}. We conclude that, to the lowest order in the current, the nucleation rate at the left boundary of the film is just
\begin{align}
\gamma\left(T,I_L\right)\approx\gamma\left(T\right)\left(1+\frac{2\pi\mathcal{P}\hbar I_L\mathcal{Q}}{ek_BT}\right).
\label{eq:gamma_V}
\end{align}

\subsection{Stability}

\label{subsec:stability}


The results of the preceding sections rely on the energy stability of skyrmion textures in the bulk of the thin film. Notice that skyrmion solutions are found in the Heisenberg Hamiltonian,\cite{Belavin_Polyakov} whose energy --$E_0=4\pi \mathcal{A}|\mathcal{Q}|$, $\mathcal{A}$ being the magnetic stiffness-- does not depend on their size given the scale invariance of the exchange interaction in 2D. Thus, there is no energy barrier that prevents their collapse into atomic-size defects. A Dzyaloshinskii-Moriya coupling term\cite{DM} selects one of the chiralities, introducing a characteristic length scale below which the shrinking of the texture is energetically penalized.

For the sake of concretion, we consider the following energy functional:\cite{Bogdanov-Yablonskii}
\begin{align}
\mathcal{U}\left[\mathbf{n}\left(\mathbf{r}\right)\right]=\int d^2\mathbf{r}\left(\mathcal{A}\text{ }\partial_i\mathbf{n}\cdot\partial_i\mathbf{n}-\kappa\text{ } n_z^2\right)/2+\mathcal{H}_{DM},
\label{eq:U}
\end{align}
where $\kappa>0$ is an easy-plane magnetic anisotropy and the last term is an interfacial Dzyaloshinskii-Moriya interaction of the form
\begin{align}
\mathcal{H}_{DM}=\mathcal{D}\int d^2\mathbf{r}\left(n_z\nabla\cdot\mathbf{n}-\mathbf{n}\cdot\nabla n_z\right).
\label{eq:DM}
\end{align}
This functional is general for thin films with $C_{nv}$ point group symmetry. The mirror symmetry $z\rightarrow-z$ is expressly broken by the last term, which can be induced by proximity with a metallic substrate with large spin-orbit coupling.


The magnetic anisotropy makes the spins to be oriented along the $\hat{z}$ axis in the ground state, whereas the Dzyaloshinskii-Moriya term stabilizes skyrmion textures connecting the two possible orientations. We consider minimal energy solutions on top of the ordered ground state pointing along the positive $\hat{z}$ axis. From now on, we assume $\mathcal{D}>0$.\cite{noteD} If we write the order parameter as \begin{align*}
\mathbf{n}=\left(\sqrt{1-n_z^2}\cos\phi,\sqrt{1-n_z^2}\sin\phi,n_z\right),
\end{align*}
then the functional in Eq.~\eqref{eq:U} hosts solutions of the form $n_z\left(\mathbf{r}\right)=n_z\left(r\right)$, $\phi=\varphi$, where $\left(\varphi,r\right)$ are polar coordinates with respect to the center of mass of the skyrmion. The polarization $n_z\left(r\right)$ satisfies the following equation:
\begin{gather}
\nonumber
\frac{1}{1-n_z^2}\left(\partial_r^2+\frac{1}{r}\partial_r\right) n_z+\frac{n_z}{\left(1-n_z^2\right)^2}\left(\partial_r n_z\right)^2
\\
+n_z\left(\frac{1}{\lambda_{\kappa}^2}+\frac{1}{r^2}\right)-\frac{2}{r\lambda_{\mathcal{D}}}\sqrt{1-n_z^2}=0,
\label{eq:Euler}
\end{gather}
with boundary conditions $n_z\left(r\rightarrow 0\right)=-1$, $n_z\left(r\rightarrow \infty\right)=1$. We have introduced length scales associated with the anisotropy and Dzyaloshinskii-Moriya interactions,\begin{gather*}
\lambda_{\kappa}\equiv\sqrt{\frac{A}{\kappa}},\\
\lambda_{\mathcal{D}}\equiv\frac{\mathcal{A}}{\mathcal{D}}.
\end{gather*}
The skyrmion charge of these solutions is 
\begin{align}
\label{eq:Q1}
\mathcal{Q}=\frac{1}{4\pi}\int_0^{2\pi} d\varphi\text{ }\partial_{\varphi}\phi\int_0^{\infty} dr \text{ }\partial_r n_z=1,
\end{align}
as it is deduced from the definition in Eq.~\eqref{eq:skyrmion_charge}.

Finding exact solutions of Eq.~\eqref{eq:Euler} is a difficult task.\cite{sol} Inspired by numerical studies,\cite{sky_numerics} we employ instead a hard cut-off variational ansatz of the form\begin{align*}
n_z\left(r\right)=\begin{cases}
\cos\left[\pi\left(1-\frac{r}{R}\right)\right]&\text{if }r\leq R,\\
1&\text{if }r> R.
\end{cases}
\end{align*}
Here $R$ represents the radius of the skyrmion, to be determined by minimizing its energy, $E_{sk}\left(R\right)$. By plugging this ansatz into Eq.~\eqref{eq:U} we obtain\begin{align}
\frac{E_{sk}\left(R\right)}{4\pi\mathcal{A}}=C-\frac{\pi}{4}\frac{R}{\lambda_{\mathcal{D}}}+\frac{1}{16}\frac{R^2}{\lambda_{\kappa}^2},
\label{eq:energetics}
\end{align}
where $C$ is a numerical constant of order 1.\cite{noteC} $E_{sk}\left(R\right)$ is plotted in Fig.~\ref{fig:sky_ener}. The value of $R$ that minimizes the energy is just\begin{align}
R_0=2\pi\frac{\lambda_{\kappa}^2}{\lambda_{\mathcal{D}}},
\end{align}
and therefore the energy of stable skyrmions\cite{stability} reduces to\begin{align}
E_0=4\pi\mathcal{A}\left(C-\frac{\pi^2\mathcal{D}^2}{4\kappa\mathcal{A}}\right).
\end{align}
The energy barrier that precludes their collapse can be estimated as\begin{align}
E_{b}=E_{sk}\left(R\sim0\right)-E_{0}=\frac{\pi^3\mathcal{D}^2}{\kappa}.
\end{align}

\begin{figure}
\includegraphics[width=1\columnwidth]{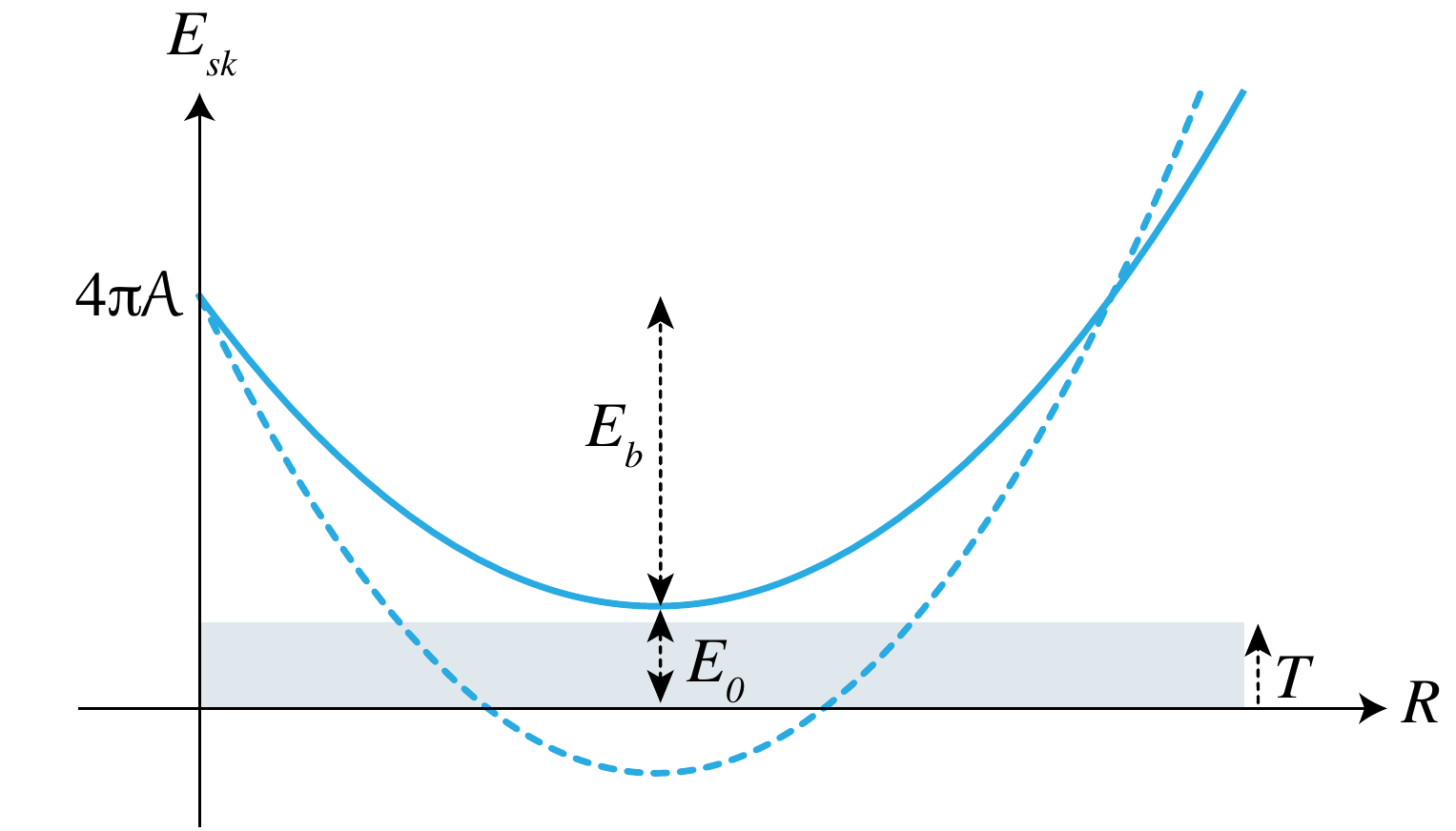}
\caption{Characteristic energy profile of a skyrmion in a thin film as a function of its radius. The continuum line corresponds to metastable excitations. The shaded region represents the temperature of the system, $T$. If the Dzyaloshinskii-Moriya interaction is very strong, the uniformly ordered state becomes unstable, as represented by the dashed line.}
\label{fig:sky_ener}
\end{figure}

\section{Topological spin drag}

\label{sec:drag}

In the two-terminal geometry depicted in Fig.~\eqref{fig:geometry}, we assume translational invariance along the transverse ($\hat{y}$) direction. According to Eq.~\eqref{eq:continuity}, the dynamics of the skyrmion charge density in the bulk of the system is given by\begin{gather}
\dot{\rho}+\text{ }\partial_xj_x=0,
\end{gather}
Then, the longitudinal skyrmion charge current in the steady-state reduces to\cite{drift}
\begin{align}
j_x=\frac{D}{L}\left(\rho_L-\rho_R\right),
\end{align}
where $\rho_{L/R}$ denotes the skyrmion charge density at the left/right terminals, and $L$ is the distance between them.

From Eqs.~\eqref{eq:reaction-rate} and \eqref{eq:gamma_V}, we can write the skyrmion charge current injected into the system from the left contact as
\begin{align}
j_L=\gamma_L\left(T\right)\left(1+\frac{2\pi\mathcal{P}\hbar I_L}{ek_BT}\right)-\bar{\gamma}_L\left(T\right)\rho_L,
\end{align}
where we are taking $\mathcal{Q}=1$ as implied by Eq.~\eqref{eq:Q1}. In the right lead, the skyrmion charge leaves the system at a rate
\begin{align}
j_R=\bar{\gamma}_{R}\left(T\right)\rho_R-\gamma_R\left(T\right).
\end{align}
The conservation of the skyrmion current implies $j_L=j_R=\frac{D}{L}\left(\rho_L-\rho_R\right)$, from which we deduce the following expression for the steady-state skyrmion charge current:
\begin{align}
j_x=\frac{2\pi\hbar\mathcal{P} I_L\rho_0}{ek_BT\left(\frac{L}{D}+\frac{1}{\bar{\gamma}_L\left(T\right)}+\frac{1}{\bar{\gamma}_R\left(T\right)}\right)}.
\end{align}

The skyrmion charge leaving the magnet sustains a voltage in the right terminal, which reads as
\begin{align}
V=-\left(\frac{2\pi\hbar\mathcal{P}}{e}\right)^2\frac{\sigma d\rho_0}{k_BT\left(\frac{L}{D}+\frac{1}{\bar{\gamma}_L\left(T\right)}+\frac{1}{\bar{\gamma}_R\left(T\right)}\right)}RI_L.
\end{align}
Here $R$ is the resistance of the metal contact. The signal is characterized by a dimensionless drag coefficient of the form\begin{align}
\mathcal{C}_d=-\frac{\sigma\mathcal{P}^2R_0^2}{R_{bulk}+R_L+R_R},
\label{eq:I_sc}
\end{align}
where $R_0\equiv\pi\hbar/e^2$ is the resistance quantum, and $R_{bulk}$, $R_{L/R}$ represent the drag resistances of the bulk and interfaces of the film, given by\begin{gather}
\label{eq:R_bulk}
R_{bulk}=\frac{k_BTL}{4e^2d\rho_0D},\\
\label{eq:R_lead}
R_{L/R}=\frac{k_BT}{4e^2d\gamma_{L/R}\left(T\right)}.
\end{gather}

\section{Discussion}
\label{sec:discussion}

For lengths much larger than $D/\gamma_{L,R}$, the boundary resistances can be neglected and the drag coefficient is limited by the mobility of the skyrmions in the bulk of the system. The expression for $\mathcal{C}_d$ reduces to Eq.~\eqref{eq:Cd}. We may distinguish 3 different regimes then:
\begin{enumerate}[(a)]
\item $T\lesssim E_0\ll E_b$: skyrmion gas,
\item $E_0<T\ll E_b$: skyrmion liquid,
\item $E_0<0$: skyrmion crystal.
\end{enumerate}

Regime (a) is depicted in Fig.~\ref{fig:sky_ener} (continuum line). A finite density of skyrmions sustains the drag signal. Its concentration is low enough that particle-particle interactions can be safely ignored. The results in Eqs.~\eqref{eq:I_sc}-\eqref{eq:R_bulk} are strictly valid as long as $E_0\ll E_b$, which guarantees the topological protection of the skyrmion charge. For films with interfacial Dzyaloshinskii-Moriya interaction and easy-axis anisotropy, this condition is fulfilled if
\begin{align*}
\frac{\pi^2 \mathcal{D}^2}{4\kappa\mathcal{A}}\lesssim 1.
\end{align*}

As the temperature increases, the concentration of skyrmions grows and eventually the dilute approximation breaks down. Many-body effects become relevant, making the skyrmion ensemble stiffer and lowering the mobility. The drag coefficient is expected to decrease as a function of temperature, manifesting the crossover to a liquid state, regime (b). However, the algebraical decay $\mathcal{C}_d\propto-1/L$ survives as long as $T\ll E_b$.

On the contrary, if relativistic interactions are very strong, the uniformly ordered ground state becomes unstable. That is the situation depicted by the dashed line in Fig.~\ref{fig:sky_ener}, corresponding to regime (c). The system undergoes a transition towards a helically ordered state or, in the presence of a magnetic field, a crystal phase in which skyrmions form a rigid lattice. The mobility drops to $0$. Nevertheless, a residual spin drag signal may survive mediated by vibrational modes of the skyrmion lattice,\cite{crystal} whose contribution is beyond the scope of our theory.

In summary, a long-ranged spin drag can be stablished in magnetic thin films in between metallic contacts, sustained by a steady-state skyrmion charge current. The injection of skyrmion charge and its conversion to an electronic current stems from the spin-transfer torque and the reciprocal spin pumping effect in the adiabatic limit, i. e. when spin relaxation at the interface is negligible. The drag coefficient decays algebraically with the distance between terminals, manifesting the topological protection of skyrmions. These ideas can be explored in thin films of chiral magnets or systems with interfacial Dzyaloshinskii-Moriya interactions. In the former case, films of the insulating ferrimagnet Cu$_2$OSeO$_3$\cite{Cu2OSeO3_film} are particularly appealing for our proposal. The different behaviors of the drag coefficient as a function of temperature and magnetic field provide a way to study the phase diagram of these materials.

\section*{Acknowledgements}

This work was supported by the U.S. Department of Energy, Office of Basic Energy Sciences under Award No. DE-SC0012190 (H.O.), and by the Army Research Office under Contract No. 911NF-14-1-0016 (S.K.K.).


\begin{thebibliography}{60}
\bibitem{attempt} The attempt frequency is related to the second derivative of the free-energy of the film with respect to the order parameter around an equilibrium solution.\cite{reaction-rate} Since the torque in Eq.~\eqref{eq:Berger} couples linearly to the order parameter as an external field $\mathbf{h}_{\tau}=\boldsymbol{\tau}\times\mathbf{n}$, we do not expect $\nu\left(T\right)$ to depend on the current, and therefore we can take its value at equilibrium.
\bibitem{noteD} In the case of $\mathcal{D}<0$, solutions with $\phi=\varphi+\pi$ are stabilized instead; the rest of the discussion follows.
\bibitem{noteC} $C=\frac{\pi^3}{2}+\frac{\pi}{2}\left(\Gamma-Ci\left(2\pi\right)+\ln 2\pi\right)$, where $\Gamma$ is Euler's constant and $Ci$ is the cosine integral function.
\bibitem{drift} In the bulk, only the conservative force $\vec{F}=-\frac{\delta \mathcal{U}}{\delta \vec{\mathcal{R}}}$ acts on the skyrmion. In the dilute limit we can neglect many-body effects, and therefore $\vec{F}=0$ as consequence of translational invariance. Only stochastic forces of thermal origin act on the skyrmion. Hence, $\vec{v}=0$ on avearge, and only the diffusive contribution to the skyrmion charge current is considered.
\end{thebibliography}
\end{document}